\documentclass[12pt]{iopart}

\pdfoutput=1

\usepackage{graphicx,subfigure}  

\newcommand{\intd}{\, \mathrm{d}}
\newcommand{\E }{E\, }
\newcommand{\phihat}{\hat{\phi}}

\usepackage{iopams}
  
\begin{document}

\title{State estimation of long-range correlated non-equilibrium systems: media estimation}

\author{Otto Pulkkinen}

\address{Fachrichtung Theoretische Physik,
Universit\"at des Saarlandes, 
 66123 Saarbr\"ucken, Germany}
\ead{otto.pulkkinen@lusi.uni-sb.de}
\begin{abstract}
Non-equilibrium systems have long-ranged spatial correlations even far away from critical points. This implies that the likelihoods of spatial steady state profiles of physical observables are nonlocal functionals. In this letter, it is shown that these properties are essential to a successful analysis of a functional level inverse problem, in which a macroscopic non-equilibrium fluctuation field is estimated from limited but spatially scattered information. To exemplify this, we dilute an out-of-equilibrium fluid flowing through random media with a marker, which can be observed in an experiment. We see that the hidden variables describing the random environment result in spatial long-range correlations in the marker signal. Two types of statistical estimators for the structure of the underlying media are then constructed: a linear estimator provides unbiased and asymptotically precise information on the particle density profiles, but yields negative estimates for the effective resistances of the media in some cases. A nonlinear, maximum likelihood estimator, on the other hand, results in a faithful media structure, but has a small bias. These two approaches complement each other. Finally, estimation of non-equilibrium fluctuation fields evolving in time is discussed.

\end{abstract}

\maketitle

\section{Introduction}

Interacting non-equilibrium systems exhibit spatial long-range correlations, in which the equal-time autocorrelation functions of physical observables show slower than exponential decay as a function of the distance between points of measurement. Presence of conservation laws, spatial anisotropies and lack of detailed balance are typical requirements for their emergence \cite{Dorfman:1994}. In some fluids, the couplings between hydrodynamic fields is the dominant cause for the correlations \cite{OrtizdeZarate:2004}. In contrast to thermodynamic equilibrium, proximity of a phase boundary is not needed.      

The long-range correlations show up in the fluctuations of observables on length-scales comparable to the system size. The statistics of these fluctuations are captured by the likelihoods of macroscopic profiles, which often satisfy a large deviation principle. In those cases, the likelihood is characterized by a large deviation functional, a generalization of the equilibrium concept of free energy, which measures the macroscopic fluctuations in terms of exponentially small probabilities \cite{Bertini:2002}. As an example, the closeness of a stationary but fluctuating fluid density field $\phi$ to an arbitrary function $f$ in a volume $V$ could be described by a large deviation functional $\mathcal{F}$ as
\begin{equation} 
\label{LDF}
P( \phi \sim f ) \approx \e^{-V \mathcal{F}(f)}
\end{equation}
with $\mathcal{F}$ attaining its minimum at the most likely profile.

It has been shown that the non-equilibrium large deviation functionals are \emph{nonlocal}, in that the likelihood of a profile is not additive under the operation of uniting two subsystems into a larger, joint system \cite{Derrida:2002}. Instead, the probability of observing a given profile in a subdomain depends also on the form of the same profile elsewhere, {\it i.e.}\ on the global structure. This is a consequence of spatial long-range correlations.

Long-range correlations and nonlocal profile likelihoods suggest that it is possible to analyze a \emph{functional level inverse problem}: even small pieces of information on local, spontaneous fluctuations can be translated by an analytical and numerical machinery to the language of fluctuations on a global scale. Building such machinery is the topic of this letter. Working through a specific application of determining the structure of random media using marker particle data, we see how to extract useful information about the fluctuation state of a random, macroscopic field from a weak signal.          

The problem of estimating the random media structure is attractive not only because of its potential applications, but also because it offers a unique view to spatial correlations in non-equilibrium systems and their state estimation in general. First, we see that hidden information, in this case a static random environment, can bring about an effective particle-particle interaction which results in long-ranged spatial correlations. In fact, for the system under study, the correlations are entirely due to hidden variables because of a special symmetry in the dynamics. The analysis is then transparent because estimation is based on a single type of correlation. Second, these long-range correlations turn out to be of a very common form, given by a piecewise linear covariance. Such covariance functions can be found in a class of driven diffusive systems (including exclusion and KMP processes \cite{Bertini:2007}). Also the correlations induced by a coupling of temperature and velocity fluctuations in a Rayleigh-B\'enard system seem to be of the same form \cite{OrtizdeZarate:2004}. We discuss a general technique for state estimation of piecewise linearly correlated systems. 

In the following sections, we define the problem of media estimation more precisely and describe two complementary solutions. State estimation of non-equilibrium systems with time-dependent fluctuation fields is then briefly discussed.  

\section{Media estimation}

We study transport of particles through one-dimensional random media, which is modeled by random single-particle transition rates between microscopic unit cells. The size of the unit cells is determined by the media correlation length, above which the transition rates have independent statistics. We assume spatial homogeneity of the media in a statistical sense. The rate at which particles move from the microscopic cell $i$ to cell $i+1$ and vice versa, is a random number $v_i$, and these random numbers are independent and identically distributed. The system is driven out of equilibrium by a chemical potential difference at two boundary reservoirs. The transition rates $v_0$ and $v_L$ at the boundaries are set equal to unity. 
 
The stationary states of particle transport in random media were studied in \cite{Pulkkinen:2007} for a class of particle interactions. In these particle systems, the \emph{fugacity profile} is a monotone function, which can be expressed in terms of partial resistances $R_j = 1 + \sum_{i=1}^{j-1} v_i^{-1}$ as 
\begin{equation}
\label{mscopic_phi}
\phi_j =\phi_- \left(1- \frac{R_j}{1+R_L} \right) + \phi_+  \frac{R_j}{1+R_L},\quad j=1,\ldots, L,
\end{equation}
where $ \phi_-$ and  $\phi_+$ are the fugacities at the left and right boundary reservoirs, respectively. The number of particles in a cell $j$ is a function of $\phi_j$ only. Thus the structure of  stationary density profiles on a \emph{macroscopic} scale depends on the statistics of the resistances $v_i^{-1}$. For finite expected resistances, the fugacity converges to a linear, deterministic function by the law of large numbers as the number of unit cells diverges. On the other hand, for $\E v_i^{-1} = \infty$, percolation effects are important. In terms of a macroscopic spatial coordinate $x\in [0,1]$, the fugacity converges in distribution to a random function
\begin{equation}
\label{phi}
\phi(x) =\phi_- \left(1- \frac{R(x)}{R(1)} \right) + \phi_+ \frac{R(x)}{R(1)},
\end{equation}
where $R$ is an $\alpha$-stable non-decreasing L\'evy process \cite{Sato.book}. Here $\alpha +1 \in (1,2)$ is the power-law exponent for the tail of the distribution for the resistances $v_i^{-1}$. For $\alpha\lesssim 1$, the fugacity profile has many small jumps, but it typically stays close to the expected profile 
\begin{equation}
\label{phimean}
\rho(x) := \E \phi(x) =  \phi_- (1- x) + \phi_+ x.
\end{equation}
On the other hand, for $\alpha$ small, the total resistance is dictated by just a few bottlenecks, which show up as large jumps in the fugacity profile, and as plateaus between the jumps (see figure \ref{fig1} below for examples). This is captured by the non-vanishing covariance 
\begin{equation}
\label{phicov}
\!\!\!\!\!\!\!\!\!\!\!\!\!\!\!\! C(x,y) := \E \phi(x)\phi(y) -\rho(x)\rho(y) = (1-\alpha)(\phi_+ - \phi_- )^2 (x\wedge y) \left( 1 - x\vee y \right),
\end{equation}
where $\wedge$ and $\vee$ denote the minimum and maximum of two numbers, respectively. 

We consider the limit of \emph{weak particle-particle interactions}, in which case fugacity equals the particle density \cite{Pulkkinen:2007}. By equation (\ref{phicov}), the correlations in the density profile are non-exponential, and mediated by the disordered media. They also vanish in equilibrium, {\it i.e.}\ when $\phi_- = \phi_+ $. Thus static hidden variables in a non-equilibrium system can result in an effective interaction, and to long-range correlations.

To obtain an image of the underlying media, we dilute the boundary reservoirs with marker particles. Only those can be observed in an experiment. For simplicity, the fraction of markers is taken to be the same in both reservoirs, but still $\phi_- \neq \phi_+$. For a very dilute marker, the numbers of observed marker particles in disjoint sets of macroscopic size within the media are asymptotically independent and Poisson distributed with the intensity of the distribution proportional the integral of the particle density\footnote{Mathematically, this is obtained by having a fraction of markers in each reservoir scale in inverse proportion to the number of unit cells.}.
\begin{equation}
\!\!\!\!\!\!\!\! P \left( \cap_{i=1}^k \lbrace N(A_i) = m_i \rbrace \vert \phi \right) = \prod_{i=1}^k \frac{\Lambda(A_i)^{m_i}}{m_i !} \e^{-\Lambda(A_i)} , \quad  
\Lambda(A_i) = c \int_{A_i }\phi(x) \intd x\, ,
\end{equation}
where $A_i \subset [0,1]$ are disjoint macroscopic sets. In other words, a steady state snapshot of marker particles is a Cox (or doubly stochastic Poisson point) process directed by a random measure $\Lambda$ with density $c\, \phi$ \cite{DaleyVere-Jones.books, Grandell.lnm}. In the following, we take $c=1$.

Given the relation (\ref{mscopic_phi}) between the the partial resistances and the particle density, we see that
\begin{equation}
\label{relres}
r(a,b) = r(a,b; \phi)  = \frac{  \phi (b) - \phi (a) }{\phi_+ - \phi_-}  
\end{equation}
is the resistance of a macroscopic interval $(a,b]$ relative to the total resistance of the media (which can be determined from a flow experiment).  In order to infer this quantity for any $a$ and $b$ from a snapshot of the marker particles, {\it i.e.}\ the Cox process data at hand, we need a reliable estimator for the density profile $\phi$. Bayesian analysis shows the \emph{minimum mean square error estimator} (MMSE estimator) of $\phi$ given markers at locations $(x_1,\ldots, x_n)$ is \cite{Grandell.lnm,Karr:1983}
\begin{equation}
\label{MMSEE}
\phihat(x) = \frac{\E \phi(x)\, \e^{-\int_{[0,1]} \phi(y)\intd y} \prod_{i=1}^{n} \phi(x_i) }{ \E \e^{-\int_{[0,1]} \phi(y)\intd y} \prod_{i=1}^{n} \phi(x_i)}.
\end{equation}
An obvious choice for the estimator of the relative resistance would then be
\begin{equation}
\hat{r}(a,b) = r(a,b; \hat{\phi}).
\end{equation}
However, expectations in expression (\ref{MMSEE}) are not analytically tractable to the author's knowledge (even the single-point moments $\E \phi(x)^n$ have quite complicated formulae \cite{Pulkkinen:2007}), and therefore good approximative estimators are needed.

\subsection{Linear estimation}

The MMSE estimator (\ref{MMSEE}) is in general a nonlinear function of the observations. In particular, it lacks additivity under inclusion of new information. Nonlinear estimation can be computationally costly because the estimator has to be recalculated completely each time new data becomes available. We next consider linear estimators of the form     
\begin{equation}
\label{LE}
\phihat_{\mathrm{L}} (x) = \int_{[0,1]} \! k(x,y)  \intd N(y), 
\end{equation}
where $N$ is the counting measure for marker particles in a snapshot ({\it i.e.}\ a Poisson random measure with random directing intensity measure $\Lambda(A) = \int_A \phi(x)\, \intd x$). In particular, $\int_{[0,1]} k(x,y) \intd N(y) = \sum_{i=1}^n k(x,x_i)$ for $n$ observed particles at locations $x_i$. The deterministic kernel $k$ translates each observed particle to the language of global density fluctuations separately.

The original construction of a \emph{MMSE linear estimator} for a general Cox process with a directing density is due to Grandell \cite{Grandell:1971}. An uncomplicated derivation uses the fact that the space in which the estimator errors are measured, the space of square integrable functions $L_2(P)$, is a  Hilbert space. The trick is that the MMSE linear estimator is a projection to the linear subspace spanned by functionals of the form $ \int_{[0,1]} f(y)  \intd N(y)$  \cite{Karr.book}. Thus the error function $\phi- \phihat_{\mathrm{L}}$ has to be orthogonal to every such functional: 
\begin{equation}
\label{Orthog}
\E \Big\lbrace \lbrack \phi(x) - \phihat_{\mathrm{L}} (x) \rbrack \int_{[0,1]} \!  f(y) \intd N(y)  \Big\rbrace = 0.
\end{equation}   
A straightforward calculation\footnote{Observe that $\E \intd N(x) \intd N(y) = \left( \delta (x- y) \E \phi(x) + \E \phi(x) \phi(y) \right) \intd x \intd y\, $.} shows that the orthogonality relation is solved by 
\begin{equation}
\label{MMSELE} 
\phihat_{\mathrm{L}} (x) =\rho(x) + \int_{[0,1]} \! K(x,y) \lbrack \intd N(y) - \rho(y)\intd y \rbrack
\end{equation}
if the kernel $K$ satisfies the integral equation   
\begin{equation}
\label{KEq}
K(x,y) \rho(y) + \int_{[0,1]} \! K(x,z) C(z,y) \intd z = C(x,y).
\end{equation}
A remarkable feature of the result (\ref{MMSELE}),(\ref{KEq}) is that only the mean $\rho(x)$ and the covariance $C(x,y)$ of the directing density are used in the construction. The information on higher moments and correlations is neglected. In this sense, linear estimation is a \emph{mean field approximation} reminiscent of Gaussian approximations.

For media estimation with the mean and covariance of the particle density given by equations (\ref{phimean}) and (\ref{phicov}), equation (\ref{KEq}) for the kernel becomes a coupled pair of integral equations because the covariance function $C(x,y)$ is piecewise defined. However, due to piecewise linearity of $C(x,y)$, differentiation of these equations twice with respect to $y$ leads to two decoupled second order differential equations. Consequently, the linear media estimation has an explicit solution:  
\begin{equation}
\!\!\!\!\!\!\!\!\!\!\!\!\!\!\!\!\!\!\!\!\!\!\!\! K (x,y) = \frac{ h(x,y) C(x,x )}{ h(x,x) \rho(x)   + \int_0^1 h(x,z) C (z,x) \intd z }\, , \quad h(x,y)=
g_- (x\wedge y) g_+ (x\vee y)\, ,
\end{equation}
where the auxiliary functions $g_+$ and $g_-$ are given in terms of modified Bessel functions $I_1$ and $K_1$ (see \cite{Abramowitz.book} for definition and properties) as
\begin{equation}
\!\!\!\!\!\!\!\!\!\!\!\!\!\!\!\!\!\!\!\!\!\!\!\!\!\!\!\!\!\!\!\! g_{\stackrel{+}{-}}(x) = \frac{2}{\sqrt{\rho(x)}} \bigg\lbrace I_1 \left( 2\sqrt{ (1-\alpha)\rho(x)} \right) - \frac{I_1 \big( 2\sqrt{(1-\alpha)\phi_{\stackrel{+}{-}}} \big) }{ K_1 \big( 2\sqrt{(1-\alpha)\phi_{\stackrel{+}{-}}}\big) }  K_1 \left(2\sqrt{(1-\alpha)\rho(x)} \right) \bigg\rbrace.
\end{equation}

\begin{figure}[tb]
\includegraphics[width=1.\textwidth]{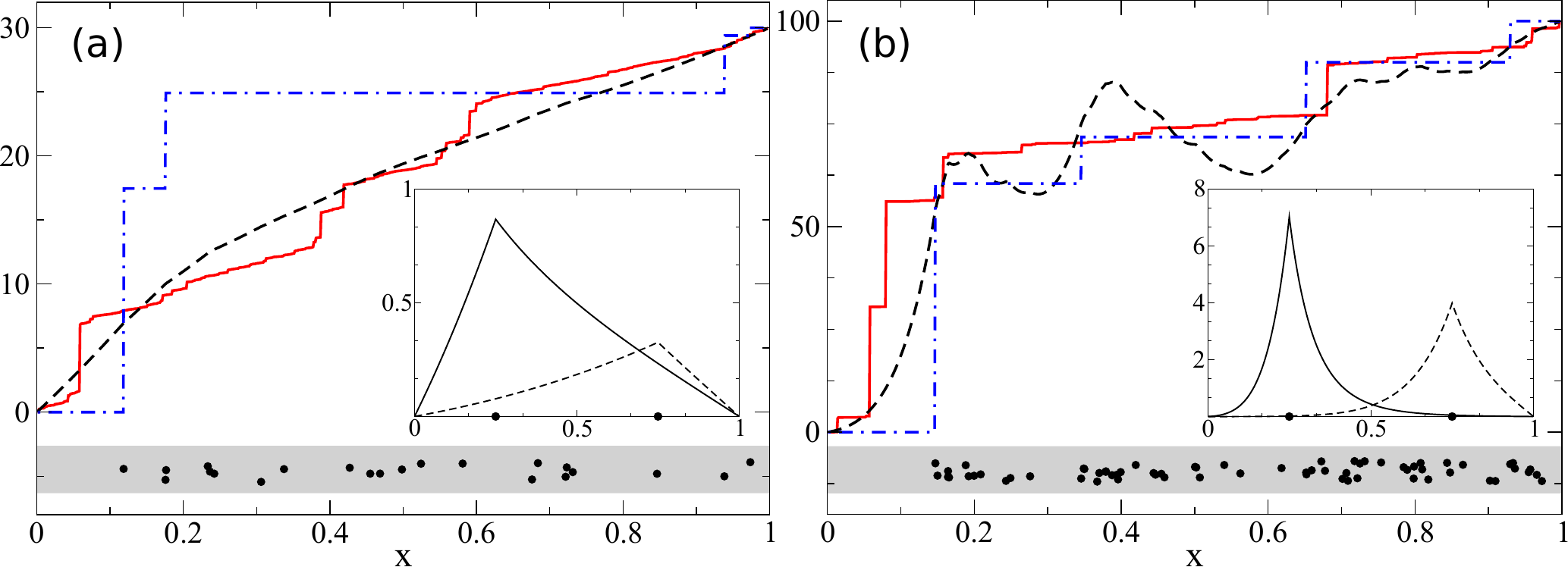}
\caption{\label{fig1} Linear (dashed) and maximum likelihood (dash-dotted) estimators for simulated marker density profiles (solid line) at (a) $\alpha=0.95$, $\phi_+ =30$, and (b) $\alpha=0.5$, $\phi_+ =100$. In both cases, $\phi_- =0$. The black circles on shaded background show the observed marker particles. Insets: the kernel functions $K(x,0.25)$ (solid line) and $K(x, 0.75)$ (dashed). }
\end{figure} 

Figure \ref{fig1} shows the linear estimator for density profiles obtained from simulations of $\alpha$-stable processes for two disorder strengths $\alpha$. In both cases, the insets show the kernel $K$, {\it i.e.}\ the effect of a single particle. As the true density profiles get closer to the expected profile as $\alpha\to 1$, also the effect a single observed particle on the estimator gets smaller, and the function $K(x,y)$ thus gets less peaked around $y$. This is due to \emph{unbiasedness} of the linear estimator, $\E \phihat_L(x) =\rho(x)$. Remarkable is also the asymmetry of the peaks, which is a consequence of the boundary conditions imposed on the density profiles.  

Figure \ref{fig1} (b) shows that the linear estimator yields non-monotone density profiles, which lead to negative resistance estimates $\hat{r}_{\mathrm{L}} (a,b ) = r(a,b;  \phihat_{\mathrm{L}} )$, defined through (\ref{relres}). The problem is not serious at large marker densities: for $0 \leq \phi_-  : = \sigma_- \gamma < \sigma_+ \gamma =: \phi_+$,
\begin{equation}
\!\!\!\!\!\!\!\!\!\!\!\!\!\!\!\!\!\!\!\!\!\!\!\!\!\!\!\!\!\!\!\!\!\!\sqrt{\gamma}\, \E \! \left \lbrack \left( \hat{r}_{\mathrm{L}} (a,b) - r(a,b) \right)^2 \right\rbrack \to \frac{ \sqrt{1-\alpha} \left( \sqrt{\sigma_+ b + \sigma_- (1-b) } + \sqrt{\sigma_+ a + \sigma_- (1-a) } \right)}{ 2(\sigma_+ -\sigma_-) } 
\end{equation}
as $\gamma \to \infty$, in that the estimation error vanishes as the inverse square root of the marker density. The proof is based on the previous Hilbert space techniques, in combination with asymptotic formulae for the modified Bessel functions.    

The constrained optimization problem of finding a linear \emph{monotone} estimator seems much more difficult to solve than the unconstrained one. However, the negative resistances can also be avoided by truncating the negative values of the linear resistance estimator $\hat{r}_{\mathrm{L}}$. This also reduces the estimation error. Alternatively, one can inspect the increments of the best monotone approximation to $ \phihat_{\mathrm{L}}$ in the uniform topology \cite{Ubhaya:1975}, 
\begin{equation}
\phihat_{\mathrm{mon,L}}(x) = \frac{1}{2} \left( \sup_{0\leq y\leq x}  \xi(y) +  \inf_{x\leq z\leq 1}  \xi(z) \right), \quad \xi(x) =  \phi_- \vee \phihat_{\mathrm{L}}(x) \wedge \phi_+.
\end{equation} 
Next we introduce a monotone estimator that complements the linear one.

\subsection{Maximum likelihood estimator}

In order to find an inherently monotone estimator, we neglect the information that we have on the statistics of the density profile increments and look for a \emph{non-decreasing} function that maximizes the likelihood function
\begin{equation}
\label{ML}
L(\phi, (x_i)) = \e^{-\int_{[0,1]} \phi(x) \intd x} \prod_{i=1}^n \phi( x_i ),
\end{equation}
under the assumption that $\phi_- \leq \phi(x_1) \leq \ldots \leq \phi(x_n) \leq \phi_+$.
The problem has been discussed for non-negative functions by numerous authors (see {\it e.g.}\ \cite{ Brunk:1955,Boswell:1966}). However, a heuristic construction is hard to find in the literature, and since we have to include the extra condition $\phi_- \leq \phi(x_1)$, we provide a derivation. 

Suppose one fixes the values $\psi_i = \phi(x_i)$ in expression (\ref{ML}). Then the likelihood maximizing function is the one that minimizes the integral in the same expression, which is nothing but the lower step function (the subscript ML stands for maximum likelihood)
\begin{equation}
\label{MLE1}
\phihat_{\mathrm{ML}}(x) = 
\left\{ \begin{array}{ll}
\phi_-  & \textrm{for}\  0 = x_0 \leq x \leq x_1 \, ,\\
\phi_- \vee \psi_i \wedge \phi_+ & \textrm{for}  \ x_i \leq x \leq x_{i+1},\quad i=1\ldots,n \, , \\
\phi_+ & \textrm{for}\  x =x_{n+1} =1\, . 
\end{array} \right. 
\end{equation}
The true problem is to find the optimal values for $\psi_i$. By inspecting the log-likelihood 
\begin{equation}
\log L( \phi, (x_i) ) = - \log \phi_- +\sum_{i=0}^{n} \lbrack \log \psi_i  - \psi_{i} \Delta x_i \rbrack, 
\end{equation}
where $\Delta x_i= x_{i+1}-x_i$, we see that if $\Delta x_{i-1} \geq \Delta x_{i}$ for every $i=1,\ldots,n$, the optimal choices are $\psi_i = 1/\Delta x_i$. 

Suppose that the required monotonicity of the gap sequence $(\Delta x_i)$ is broken at particle $i$, in that $\Delta x_{i-1} < \Delta x_{i}$. Clearly one needs to level the values $\psi_{i-1}$ and $\psi_i$ in such a way that the sequence is monotone again; $\psi_{i-1,i}=\psi_i = \psi_{i-1}$. This common value appears as $\log \psi_{i-1,i}^2 - \psi_{i-1,i} (\Delta x_{i-1} + \Delta x_{i} )$ in the log-likelihood, so the new optimizer reads $\psi_{i-1,i} = 2/(\Delta x_{i-1} + \Delta x_{i} )$. If after that $\psi_{j-2} \leq \psi_{i-1,i} \leq \psi_{j+1}$ is violated, one makes the same adjustment for a necessary number of values to the left and to the right from the particle at $x_i$. The log-likelihood involving adjustments of $\psi_j$ and $\psi_k$ with $j\leq i \leq k$ has a term $\log \psi_{i}^{k+1-j} - \psi_{i} ( \Delta x_j + \ldots + \Delta x_{k} )$, which leads to an optimal density $\psi_i = (k+1-j)/(x_{k+1}-x_j)$ on that plateau. The condition that the adjustment process ends at index $j$ at left and at $k$ at right is 
\begin{equation}
\!\!\!\!\!\!\!\!\!\!\!\!\!\!\!\!\!\!\!\!\!\!\!\! \frac{1}{\Delta x_{j-1}} \leq  \frac{k+1-j}{x_{k+1}-x_j} \leq \frac{1}{\Delta x_{k+1}} \Leftrightarrow  \frac{k+1-(j-1)}{x_{k+1}-x_{j-1}} \leq  \frac{k+1-j}{x_{k+1}-x_j} \leq \frac{k+2-j}{x_{k+2}-x_j} \, ,
\end{equation}
and hence
\begin{equation}
\label{MLE2}
\psi_i = \max_{1\leq j \leq i} \min_{i\leq k \leq n} \frac{k+1-j}{x_{k+1}-x_j}\, .
\end{equation}

The maximum likelihood (ML) estimator, given by (\ref{MLE1}) and (\ref{MLE2}), is plotted in figure \ref{fig1}. The estimated density profiles consist of discontinuities at marker particle positions and of plateaus in between them. This makes the ML estimator flexible as compared to the linear estimator and therefore particularly suited for estimation at small values of $\alpha$. For the same reason, the corresponding relative resistance measure does not have a continuous density as in linear estimation, but a spike train just like the original media. 

The lack of information on the disorder strength $\alpha$, however, poses some problems at low marker densities. Especially for $\alpha$ close to one, the estimated profiles are typically far away from the expected profile $\rho(x)$, and the ML estimator is outperformed by the linear estimator (see figure \ref{fig1} (a)). 

Contrary to the linear estimator for the density, the ML estimator is \emph{biased}. It easily underestimates the density at very low marker densities because in absence of marker particle observations, $\phihat_{\mathrm{ML}}(x) = \phi_-$ for $x<1$.  

\section{Discussion}

We have shown that long-range range correlations present in non-equilibrium systems can be used to extract information on an underlying fluctuation field even from very limited information. We applied the methods from statistical state estimation theory to particle transport in disordered media and estimated the media structure from a dilute marker signal. Linear estimation turned out to deliver unbiased but not necessarily positive estimates for the effective resistances of the media. This could be remedied by truncation or by monotonizing the density estimator. On the other hand, a nonlinear, maximum likelihood estimator yielded positive resistances but has an inherent bias, which is severe only at very small marker densities. Although quantitative results about the asymptotic error of the estimates could be derived in the linear case only, numerical simulations indicate fast convergence of the ML estimator.

The state-estimation of a non-equilibrium system from marker data was based on the assumption that the marker particles in a stationary flow are described by a Cox process. Furthermore, the directing density of the process was taken to be of the same form as the total particle density. This assumption can be broken in systems with steric effects, such as particle-particle exclusion, in which case one has to be able to tell, from which reservoir the particles originated from. 

In driven systems without quenched disorder, the long-range correlations are often weak, in that the amplitude of the correlations decays as a function of the system size. It is only the macroscopic correlations of a suitably rescaled fluctuation field that persist in the scaling limit \cite{Spohn:1983, Bertini:2007}. This is in contrast to non-vanishing correlations in the density profiles in random media. Probably a stronger marker signal is required for a successful estimation in case of weak correlations. 

The asymptotic correlations in a class of driven diffusive systems are of the same form as in the media estimation problem, apart from the system size dependent amplitude \cite{Bertini:2007}. Thus one can apply the linear estimation machinery to state estimation of large but finite systems in this class. Another approach is through microscopic solutions and generating functions such as used in the analysis of large deviations \cite{Derrida:2002}. Pursuing these themes further will give information on the feasibility of state estimation for general non-equilibrium systems.    

\ack
The author would like to thank Damien Simon for a discussion. This study was funded from a DFG grant BE 2478/2-1 and by the DFG Research Training Group GRK 1276.


\bibliographystyle{./unsrt}

\vfill

\end{document}